# Superhumps and grazing eclipses in the dwarf nova BG Arietis

Jeremy Shears, David Boyd, Tut Campbell, Franz-Josef Hambsch, Enrique de Miguel, Ian Miller, Etienne Morelle, George Roberts, Richard Sabo and Bart Staels

## Abstract

We report unfiltered photometry of BG Arietis (= SDSS J015151.87+140047.2) in 2009 and 2010 which shows the presence of superhumps with peak-to-peak amplitude of up to 0.36 magnitudes showing this to be an SU UMa type dwarf nova. The outburst amplitude was 5.1 magnitudes above a mean quiescence magnitude of 19.9. The 2010 event lasted at least 20 days during which we observed a precursor outburst which was immediately followed by a superoutburst. Modulations were observed during the rise to superoutburst with a period approximately half $P_{orb}$. The mean superhump period during the plateau phase was $P_{sh} = 0.0849(6)$ d, representing a superhump period excess $\varepsilon = 0.030(7)$. We also observed small eclipses of depth 0.06 to 0.13 magnitude and  FWHM duration of ~12 mins ($\Delta\Phi_{\frac{1}{2}} = 0.10$), which we interpret as grazing eclipses of the accretion disc. The 2009 superoutburst was less well observed, especially in the early stages, so we cannot tell whether there was a precursor outburst. The mean superhump period during the plateau was $P_{sh} = 0.0851(4)$ d, but towards the end of the plateau it suddenly decreased to $P_{sh} = 0.0845(11)$ d. Small eclipses were also observed.

## History of BG Arietis

BG Ari was first reported as an eruptive object (PG 0149+137), possibly a supernova, in the Palomar-Green survey with B= 15.7 and having a UV-excess (1) (2). Szkody *et al*. (3) subsequently identified it as a dwarf nova during the course of the Sloan Digital Sky Survey calling it SDSS J015151.87+140047.2. The SDSS spectrum contains double-peaked Balmer emission lines, suggesting a high orbital inclination. Dillon *et al*. (4) conducted g-band photometry at quiescence which showed a double-humped morphology with an amplitude of ~0.3 mag, yielding $P_{orb} = 0.082417(28)$ d.

The Catalina Real-Time Transient Survey (CRTS) shows that BG Ari varies between V~ 18.4 and 21.2 at quiescence, with a mean of V= 19.9 (5). Two outbursts are evident in the data: one in 2007Januay reaching V=15.5 and one in 2009 September reaching V= 14.8. The latter outburst is discussed in this paper along with a further one in 2010 December.

## Photometry and analysis

Unfiltered photometry was conducted during the 2009 (55 h) and 2010 (134 h) outbursts of BG Ari using the instrumentation shown in Table 1 and according to the observation log in Table 2. Images were dark-subtracted and flat-fielded prior to being measured using differential aperture photometry relative to the AAVSO V-band sequence 4146pw (6). Given that each observer used slightly different instrumentation, including CCD cameras with different spectral responses, small systematic differences are likely to exist between observers. Where overlapping datasets were obtained during the outburst, we aligned measurements by different observers by experiment. Adjustments of up to 0.06 magnitudes were made. Heliocentric corrections were applied to all data.

**The 2009 outburst**

The outburst was first detected by CRTS at V=17.8 on 2009 Aug 27 and the light curve is shown in the top panel of Figure 1, where CRTS data are shown by red data points and the authors' data are blue. Most of the first part of the outburst was missed, but we observed the later part of the plateau phase and the rapid decline towards quiescence, a total of about 16 days (if the CRTS detection was part of the same outburst, then the total outburst duration was ~30 days). The star reached magnitude 14.8 at maximum, 5.1 mag above mean quiescence.

We plot expanded views of the photometry in Figure 2 which clearly show the presence of superhumps indicating that this was a superoutburst and that BG Ari is therefore a member of the SU UMa family dwarf nova. To study the superhump behaviour, we first extracted the times of each sufficiently well-defined superhump maximum by using the Kwee and van Woerden (7) method in the *Minima v2.3* software (8). Times of 23 superhump maxima were found and are listed in Table 3a. An unweighted linear analysis of the times of maximum between HJD 2455086 and 2455092 allowed us to obtain the following superhump maximum ephemeris:

$$HJD_{max} = 2455086.5072(44) + 0.0851(4) \times E \qquad \text{Equation 1}$$

Thus the mean superhump period in this interval was $P_{sh}$ = 0.0851(4) d. The observed minus calculated (O–C) residuals for all the superhump maxima relative to the ephemeris are shown in the middle panel of Figure 1. This suggests that $P_{sh}$ was constant between HJD 2455086 and 2455092, at which point the period suddenly decreased to give a new superhump period regime. A linear analysis of the superhump maximum times between HJD 2455092 and 2455099 results in $P_{sh}$ = 0.0845(11) d for this later stage of the outburst.

The superhump amplitude was greatest at the beginning of the time resolved photometry on HJD 2455086 (0.36 mag peak-to-peak; Fig 2a), subsequently reducing in size until HJD 2455091 (Figure 1, bottom panel). The superhumps rapidly regrew on HJD 2455092 to 0.37 mag and then once again decreased. We note that the time at which the superhumps regrew coincided with the abrupt decrease in $P_{sh}$.

**The 2010 outburst**

*Superhumps*

The 2010 outburst received more dense photometry coverage than the one in 2009. The overall light curve is shown in the top panel of Figure 3 and detailed photometry in Figure 4. The outburst was detected on Dec 4 (9) when the star was still brightening. It reached a peak of mag 14.9 later that night (JD 2455535) and then started fading rapidly to below mag 15.1(Figure 4a), suggesting that this was a precursor outburst. The next night the star was again brighter (mag 14.8) and small modulations were present (Figure 4a). Full-grown superhumps were detected on JD 2455538, some 3 days after the outburst was detected (Figure 4b). This signalled the beginning of the plateau phase, which lasted some 16 days before a more rapid fade began. Unfortunately the return to quiescence was not observed. Thus the outburst lasted a total of at least 20 days, including both the precursor and the superoutburst. The amplitude of the superoutburst was 5.1 mag above mean quiescence.

As before, we measured the times of 37 superhump maxima (Table 3b) and obtained the following unweighted linear superhump ephemeris for the interval HJD 2455538 to 2455540:

$$HJD_{max} = 2455538.3144(72) + 0.0849(6) \times E \qquad \text{Equation 2}$$

Thus the mean superhump period in this interval was $P_{sh} = 0.0849(6)$ d, which is consistent with the measured value of $P_{sh}$ during the 2009 superoutburst in Equation 1.The O–C residuals for the superhump maxima relative to this ephemeris are plotted in the middle panel of Figure 3. This shows that the superhump period was constant during the majority of the plateau phase, but began to decrease towards the end. We did not observe the dramatic shortening that had been seen in 2009, possibly because we missed this stage of the outburst

The superhump amplitude was greatest on HJD 2455538 (0.36 mag). After this point, the amplitude decreased during the next 5 days to 0.19 mag on HJD 2455543, at which point they rapidly regrew to 0.30 mag before gradually declining during the rest of the plateau phase.

### Eclipses

Close inspection of the light curve revealed the presence of small dips superimposed on the superhumps, the regularity of which leads us to interpret as eclipses. The eclipses were only obvious when they coincided with superhump minimum (e.g. HJD 2455538, 2455544 and 2455550; Figure 4b, e and g). Examples of two eclipses, from HJD 2455550, are shown in Figure 5.

We measured 9 times of eclipse mimina, again using the Kwee and van Woerden method (7). These are listed in Table 4. An unweighted linear fit to these times gave the following eclipse ephemeris:

$$HJD_{min} = 2455538.3696(54) + 0.08259(19) \times E \qquad \text{Equation 3}$$

Thus the $P_{orb}$ of the system was 0.08259(19) d. Such value is consistent with the value measured by Dillon *et al.* (4), although we note the much larger error on our measurement. The O-C residuals of the eclipse minima relative to the ephemeris in Equation 3 are shown in Table 4 and Figure 6.

The eclipse depth and duration were difficult to determine accurately due to the difficulty of isolating the eclipse from the superhump profile. The eclipse depth varied between 0.06 and 0.13 mag with a mean of 0.08 mag (Table 4). We only succeeded in measuring the eclipse full width half minimum (FWHM) duration of the two eclipses: those on HJD 2455550 (Figure 5) where the duration was 12 mins. This represents a fraction of the orbital period of $\Delta\Phi_{\frac{1}{2}} = 0.10$. The short duration and shallow nature of the eclipses suggests that these are grazing eclipses of the accretion disc with the binary system being only slightly above the critical inclination for eclipses to occur. Moreover, the fact that the eclipses were only evident at some stages of the superoutburst is also constant with a marginal graze. The eclipse depth in eclipsing SU UMa systems is often strongly affected by the location of the superhump: eclipses are shallower when hump maximum coincides with eclipse, such as found, for example, in the case of DV UMa (10), IY UMa (11) and SDSS J122740.83+513925.9 (12). In such systems, since the accretion disc is elliptical and precesses, the eclipse depth varies with the precession period (which is the beat period between $P_{sh}$ and $P_{orb}$). According to the relation $1/P_{prec} = 1/P_{orb} - 1/P_{sh}$ (13), the precession period of BG Ari is about 2.8 d.

Having identified eclipses during the 2010 superoutburst of BG Ari, we revisited the 2009 data, paying particular attention to the times of superhump minimum. We found evidence of small eclipses at superhump minima on HJD 2455092 and 2455098 (Figures 2d and 2f, respectively). The presence of only very small eclipses, may be because we only caught the last part of the 2009 outburst at which point it is likely that the disc was already shrinking. Or it could be that the geometry of the disc, or the times at which we observed, was not conducive to recording an eclipse.

We note that there is no sign of an eclipse in the phase diagram presented by Dillon *et al.* (4). This is probably because the accretion disc is smallest when in quiescence, thus not even a grazing eclipse of the disc would take place.

### Superhump period excess and estimation of the binary mass ratio

Taking the measured values of $P_{orb} = 0.08242(3)$ d from Dillon et al. (4) and $P_{sh} = 0.0849(6)$ d from the 2010 superoutburst reported in the present study, allows the fractional superhump period excess $\varepsilon = (P_{sh} - P_{orb})/P_{orb}$ to be calculated as 0.030(7). This value is consistent with the range of $\varepsilon$ observed in other SU UMa dwarf novae with similar $P_{orb}$ (14). Measuring $\varepsilon$ provides a way to estimate the mass ratio, $q = M_{sec}/M_{wd}$, of a CV and following Patterson *et al.* (14) we find $q \approx 0.14$ for BG Ari.

### The precursor outburst

As discussed above, our observations of the 2010 superoutburst show the presence of a precursor outburst. Such precursor outbursts have been seen in other SU UMa dwarf novae, such as VW Hyi (15) and V342 Cam (16). We examined the photometry from the precursor on HJD 2455535 using the Lomb-Scargle algorithm in the *Peranso* software (17), but could find no significant periodic signal. By contrast, the following night (HJD 2455536) when the star had re-brightened, but before superhumps were visible, small modulations in the light curve were evident. We analysed the two photometry runs obtained on HJD 2455536 (Figure 4a) separately since a visual examination of the light curve suggested slightly different hump morphology in each run. First we performed a Lomb-Scargle period analysis on the data between HJD 2455536.3 and 2455536.6, having first subtracted the linear trend from the data. The resulting power spectrum (Figure 7a) has its highest peak at 25.70(1.18) cycles/d, or P = 0.0389(17) d. We plot a phase diagram of the photometry from this period, folded on P = 0.0389 d, in Figure 8 where modulations with a peak-to-peak amplitude of 0.04 mag can be seen. Since the measured period of these humps is approximately half $P_{orb}$ we speculate that these modulations represent orbital humps (or double humps). Orbital humps have been observed in the rise to superoutburst in other SU UMa systems (18).

By contrast, a visual inspection of the light curve from the second photometry run between HJD 2455536.7 and 2455536.9 appeared to show what might be emerging superhumps. A Lomb-Scargle power spectrum of these data had its strongest signal at 9.39(12) cycles/d, which we interpret as the superhump frequency. We need to treat this interpretation with caution due to the short time over which the observations were made. If true, this means $P_{sh} = 0.1064(14)$ d, which is about 25% longer than $P_{sh}$ measured later in the outburst. It is common in SU UMa systems for $P_{sh}$ in the early stages of an outburst to be ~1 to 4% longer than the mean $P_{sh}$ later in the outburst (19), but the value for BG Ari appears very high. Nevertheless, we note that in the case of V342 Cam, the superhump period in the early stages of the outburst was ~10% longer than the mean period later in the outburst (16). Both BG Ari and V342 Cam ($P_{orb}$ = 0.0753 d) have relatively long orbital periods among SU UMa systems, thus this behaviour might be associated with longer $P_{orb}$ systems. In reality, the evolution of superhumps in the very earliest stages of an outburst has not often been studied in detail.

It is generally agreed that superhumps are the visible manifestation of a precessing accretion disc that extends to the 3:1 resonance radius (20). The development of superhumps during the precursor outburst of BG Ari and the subsequent superoutburst is consistent with the thermal-tidal instability model (21) (TTIM), where a normal outburst results in the growth of the accretion disc to the 3:1 resonance radius, triggering the superoutburst. In this regard it is noteworthy that no eclipses were apparent prior to the development of superhumps, suggesting that the accretion disc had not yet expanded sufficiently for eclipses to occur and, presumably, for superhumps to be initiated.

**Future outbursts**

We encourage further monitoring of BG Ari for future outbursts. The interval between the 2009 and 2010 outbursts is about 15 months and the interval between the 2007 and 2009 outbursts is approximately twice this value. Thus the supercycle could be ~15 months if an outburst was missed in spring 2008 when the field was unobservable due to its proximity to the Sun. Time resolved photometry during future superoutbursts may yield further information about the nature of the eclipses, which could be used to study the variation of the size of the accretion disc during the outburst and whether the accretion discs expands to different extents from one superoutburst to another.

## Conclusions

We report unfiltered photometry of BG Ari in 2009 and 2010 which shows for the first time that this is an eclipsing SU UMa-type dwarf nova. In both cases the outburst amplitude was 5.1 magnitudes above a mean quiescence magnitude of 19.9. The 2010 event lasted at least 20 days during which we observed a precursor outburst which was immediately followed by a superoutburst. Modulations were observed during the rise to superoutburst with a period approximately half $P_{orb}$. The mean superhump period during the plateau phase was $P_{sh} = 0.0849(6)$ d, representing a superhump period excess $\varepsilon = 0.030(7)$. Superhumps with peak-to-peak amplitude of 0.36 magnitudes were observed at the beginning of the outburst, which gradually decreasing in amplitude as the outburst progressed. We also observed small eclipses of depth 0.06 to 0.13 magnitude and FWHM duration of ~12 mins ($\Delta\Phi_{\frac{1}{2}} = 0.10$), which we interpret as grazing eclipses of the accretion disc. The 2009 superoutburst, was less well observed, especially in the early stages, so we cannot tell whether there was a precursor outburst. The mean superhump period during the plateau was $P_{sh} = 0.0851(4)$ d, but towards the end of the plateau it suddenly decreased to $P_{sh} = 0.0845(11)$ d. Although eclipses were present, they were less prominent than in 2010 possibly because the outburst was only observed in detail towards the end of the plateau phase when the accretion disc would have been smaller

## Acknowledgements

The authors gratefully acknowledge the use of data from the Catalina Real-Time Transient Survey. We also used SIMBAD, operated through the Centre de Données Astronomiques (Strasbourg, France) and the NASA/Smithsonian Astrophysics Data System. We thank our referees, Prof. Boris Gänsicke (University of Warwick, UK) and Dr. Chris Lloyd (Open University, UK), for their helpful comments.

## Addresses

JS: "Pemberton", School Lane, Bunbury, Tarporley, Cheshire, CW6 9NR, UK [bunburyobservatory@hotmail.com]

DB: 5 Silver Lane, West Challow, Wantage, Oxon, OX12 9TX, UK [drsboyd@dsl.pipex.com]

TC: 7021 Whispering Pine, Harrison, AR 72601, USA [jmontecamp@yahoo.com]

FJH: Vereniging voor Sterrenkunde, CBA Mol, Belgium, AAVSO, GEOS, BAV [hambsch@telenet.be]

EdM: Departamento de Fisica Aplicada, Facultad de Ciencias Experimentales, Universidad de Huelva, 21071 Huelva, Spain; Center for Backyard Astrophysics, Observatorio del CIECEM, Parque Dunar, Matalascañas, 21760 Almonte, Huelva, Spain [demiguel@uhu.es]

IM: Furzehill House, Ilston, Swansea, SA2 7LE, UK [furzehillobservatory@hotmail.com]

EM: Lauwin-Planque Observatory, F-59553 Lauwin-Planque, France [etmor@free.fr]

GR: 2007 Cedarmont Dr., Franklin, TN 37067, USA,  [georgeroberts@comcast.net]

RS: 2336 Trailcrest Dr., Bozeman, MT 59718, USA [richard@theglobal.net]

BS: CBA Flanders, Patrick Mergan Observatory, Koningshofbaan 51, Hofstade, Aalst, Belgium [staels.bart.bvba@pandora.be]

| Observer | Telescope | CCD |
|---|---|---|
| Boyd | 0.25 m reflector | Starlight Xpress SXV-H9 |
| Campbell | 0.2 m SCT | SBIG ST-7 |
| Hambsch | 0.4 m reflector | SBIG STL-11kXM |
| de Miguel | 0.25 m reflector | QSI-516ws |
| Miller | 0.35 m SCT | Starlight Xpress SXVF-H16 |
| Morelle | 0.4 m SCT | SBIG ST-9 |
| Roberts | 0.4 m SCT | SBIG ST-8 |
| Sabo | 0.43 m reflector | SBIG STL-1001 |
| Shears | 0.28 m SCT | Starlight Xpress SXVF-H9 |
| Staels | 0.28 m SCT | Starlight Xpress MX-716 |

**Table 1: Instrumentation**

| Start time HJD | End time HJD | Duration (h) | Observer |
|---|---|---|---|
| **2009 Outburst** | | | |
| 2455086.464 | 2455086.628 | 3.9 | Miller |
| 2455087.412 | 2455087.540 | 3.1 | Hambsch |
| 2455087.451 | 2455087.687 | 5.7 | Miller |
| 2455088.377 | 2455088.567 | 4.6 | Hambsch |
| 2455088.543 | 2455088.592 | 1.2 | Miller |
| 2455089.508 | 2455089.675 | 4.0 | Miller |
| 2455091.494 | 2455091.691 | 4.7 | Miller |
| 2455092.377 | 2455092.644 | 6.4 | Hambsch |
| 2455092.450 | 2455092.691 | 5.8 | Miller |
| 2455094.377 | 2455092.620 | 5.8 | Hambsch |
| 2455095.464 | 2455095.467 | 0.1 | Miller |
| 2455098.477 | 2455098.542 | 1.6 | Miller |
| 2455099.536 | 2455099.615 | 1.9 | Miller |
| 2455101.383 | 2455101.644 | 6.3 | Hambsch |
| **2010 Outburst** | | | |
| 2455535.453 | 2455535.524 | 1.7 | Shears |
| 2455535.705 | 2455535.931 | 5.4 | Sabo |
| 2455536.385 | 2455536.560 | 4.2 | Miller |
| 2455536.702 | 2455536.817 | 2.8 | Campbell |
| 2455537.383 | 2455537.542 | 3.8 | Shears |
| 2455538.279 | 2455538.420 | 3.4 | Boyd |
| 2455538.405 | 2455538.583 | 4.3 | Miller |
| 2455538.529 | 2455538.722 | 4.6 | Sabo |
| 2455538.543 | 2455538.721 | 4.3 | Roberts |
| 2455539.237 | 2455539.491 | 6.1 | Shears |
| 2455539.384 | 2455539.526 | 3.4 | Miller |
| 2455539.484 | 2455539.731 | 5.9 | Roberts |
| 2455539.637 | 2455539.709 | 1.7 | Sabo |
| 2455540.567 | 2455540.824 | 6.2 | Sabo |
| 2455540.635 | 2455540.775 | 3.4 | Campbell |
| 2455541.328 | 2455541.512 | 4.4 | Morelle |
| 2455541.380 | 2455541.519 | 3.4 | De Miguel |
| 2455542.303 | 2455542.509 | 4.9 | Morelle |
| 2455543.266 | 2455543.527 | 6.3 | Staels |
| 2455543.293 | 2455543.376 | 2.0 | Miller |
| 2455544.336 | 2455544.503 | 4.0 | Morelle |
| 2455544.337 | 2455544.544 | 5.0 | De Miguel |
| 2455545.318 | 2455545.453 | 3.2 | De Miguel |
| 2455545.361 | 2455545.457 | 2.3 | Morelle |
| 2455549.217 | 2455549.300 | 2.0 | Morelle |
| 2455550.228 | 2455550.271 | 1.0 | Shears |
| 2455550.297 | 2455550.377 | 1.9 | Boyd |
| 2455550.547 | 2455550.749 | 4.8 | Roberts |
| 2455551.534 | 2455551.664 | 3.1 | Sabo |
| 2455552.540 | 2455552.665 | 3.0 | Sabo |
| 2455553.278 | 2455553.464 | 4.5 | Shears |
| 2455554.293 | 2455554.434 | 3.4 | Shears |
| 2455554.533 | 2455554.769 | 5.7 | Sabo |
| 2455555.339 | 2455555.479 | 3.4 | Miller |
| 2455555.533 | 2455555.715 | 4.4 | Sabo |

**Table 2: Observation log**

| Superhump cycle number | Superhump max (HJD) | Uncertainty (d) | O-C (d) | Superhump amplitude (mag) |
|---|---|---|---|---|
| 0 | 2455086.5089 | 0.0018 | 0.0017 | 0.36 |
| 1 | 2455086.5899 | 0.0006 | -0.0024 | 0.30 |
| 11 | 2455087.4418 | 0.0030 | -0.0015 | 0.27 |
| 12 | 2455087.5303 | 0.0009 | 0.0019 | 0.27 |
| 12 | 2455087.5313 | 0.0024 | 0.0029 | 0.27 |
| 13 | 2455087.6156 | 0.0006 | 0.0021 | 0.29 |
| 23 | 2455088.4629 | 0.0012 | -0.0016 | 0.29 |
| 24 | 2455088.5472 | 0.0012 | -0.0024 | 0.29 |
| 24 | 2455088.5492 | 0.0024 | -0.0001 | 0.29 |
| 36 | 2455089.5690 | 0.0003 | -0.0015 | 0.25 |
| 37 | 2455089.6554 | 0.0009 | -0.0002 | 0.24 |
| 59 | 2455091.5278 | 0.0006 | 0.0000 | 0.23 |
| 60 | 2455091.6128 | 0.0012 | -0.0001 | 0.25 |
| 70 | 2455092.4645 | 0.0006 | 0.0006 | 0.28 |
| 70 | 2455092.4623 | 0.0027 | -0.0016 | 0.31 |
| 71 | 2455092.5506 | 0.0006 | 0.0016 | 0.34 |
| 71 | 2455092.5505 | 0.0039 | 0.0015 | 0.32 |
| 72 | 2455092.6351 | 0.0006 | 0.0010 | 0.37 |
| 72 | 2455092.6342 | 0.0024 | -0.0004 | 0.37 |
| 93 | 2455094.4017 | 0.0054 | -0.0200 | 0.27 |
| 94 | 2455094.4959 | 0.0027 | -0.0109 | 0.25 |
| 95 | 2455094.5765 | 0.0021 | -0.0154 | 0.26 |
| 154 | 2455099.5654 | 0.0057 | -0.0474 | 0.05 |

**(a) 2009 Outburst**

| Superhump cycle number | Superhump max (HJD) | Uncertainty (d) | O-C (d) | Superhump amplitude (mag) |
|---|---|---|---|---|
| 0 | 2455538.3057 | 0.0006 | -0.0087 | 0.26 |
| 1 | 2455538.3928 | 0.0006 | -0.0065 | 0.26 |
| 3 | 2455538.5746 | 0.0012 | 0.0054 | 0.34 |
| 3 | 2455538.5750 | 0.0006 | 0.0058 | 0.32 |
| 3 | 2455538.5751 | 0.0003 | 0.0059 | 0.34 |
| 4 | 2455538.6572 | 0.0021 | 0.0030 | 0.35 |
| 4 | 2455538.6599 | 0.0006 | 0.0058 | 0.36 |
| 11 | 2455539.2489 | 0.0003 | 0.0002 | 0.35 |
| 12 | 2455539.3304 | 0.0003 | -0.0033 | 0.35 |
| 13 | 2455539.4151 | 0.0009 | -0.0035 | 0.33 |
| 13 | 2455539.4163 | 0.0006 | -0.0023 | 0.34 |
| 14 | 2455539.5026 | 0.0003 | -0.0009 | 0.33 |
| 14 | 2455539.5018 | 0.0003 | -0.0017 | 0.36 |
| 15 | 2455539.5850 | 0.0018 | -0.0035 | 0.35 |
| 16 | 2455539.6735 | 0.0015 | 0.0001 | 0.32 |
| 16 | 2455539.6743 | 0.0012 | 0.0004 | 0.32 |
| 28 | 2455540.6945 | 0.0006 | 0.0018 | 0.32 |
| 28 | 2455540.6939 | 0.0009 | 0.0007 | 0.34 |
| 29 | 2455540.7791 | 0.0024 | 0.0015 | 0.31 |
| 36 | 2455541.3730 | 0.0006 | 0.0008 | 0.34 |
| 37 | 2455541.4565 | 0.0006 | -0.0007 | 0.34 |
| 37 | 2455541.4564 | 0.0006 | -0.0012 | 0.33 |
| 47 | 2455542.3038 | 0.0006 | -0.0028 | 0.26 |
| 48 | 2455542.3880 | 0.0006 | -0.0035 | 0.26 |
| 49 | 2455542.4723 | 0.0009 | -0.0041 | 0.25 |
| 59 | 2455543.3230 | 0.0012 | -0.0028 | 0.22 |
| 59 | 2455543.3232 | 0.0009 | -0.0031 | 0.20 |
| 60 | 2455543.4097 | 0.0024 | -0.0011 | 0.25 |
| 61 | 2455543.4946 | 0.0027 | -0.0011 | 0.19 |
| 72 | 2455544.4307 | 0.0009 | 0.0006 | 0.29 |
| 72 | 2455544.4312 | 0.0015 | 0.0007 | 0.30 |
| 83 | 2455545.3627 | 0.0036 | -0.0017 | ND |
| 129 | 2455549.2703 | 0.0015 | -0.0013 | 0.22 |
| 145 | 2455550.6271 | 0.0006 | -0.0036 | 0.19 |
| 146 | 2455550.7121 | 0.0006 | -0.0035 | 0.19 |
| 168 | 2455552.5719 | 0.0027 | -0.0124 | 0.11 |
| 192 | 2455554.6125 | 0.0005 | -0.0104 | ND |

**(b) 2010 Outburst**

## Table 3: Superhump maximum times

ND: not determined

| Eclipse number | Eclipse minimum (HJD) | Uncertainty (d) | O-C (d) | Eclipse depth (mag) |
|---|---|---|---|---|
| 0 | 2455538.3652 | 0.0030 | -0.0045 | 0.1 |
| 1 | 2455538.4519 | 0.0045 | -0.0004 | ND |
| 2 | 2455538.5351 | 0.0033 | 0.0002 | 0.1 |
| 3 | 2455538.6163 | 0.0039 | -0.0012 | 0.08 |
| 4 | 2455538.7038 | 0.0045 | 0.0037 | 0.06 |
| 73 | 2455544.3984 | 0.0036 | -0.0004 | 0.07 |
| 74 | 2455544.4855 | 0.0036 | 0.0041 | 0.06 |
| 148 | 2455550.5922 | 0.0033 | -0.0008 | 0.08 |
| 149 | 2455550.6745 | 0.0039 | -0.0011 | 0.13 |

## Table 4: Eclipse times (2010 outburst)

ND: not determined

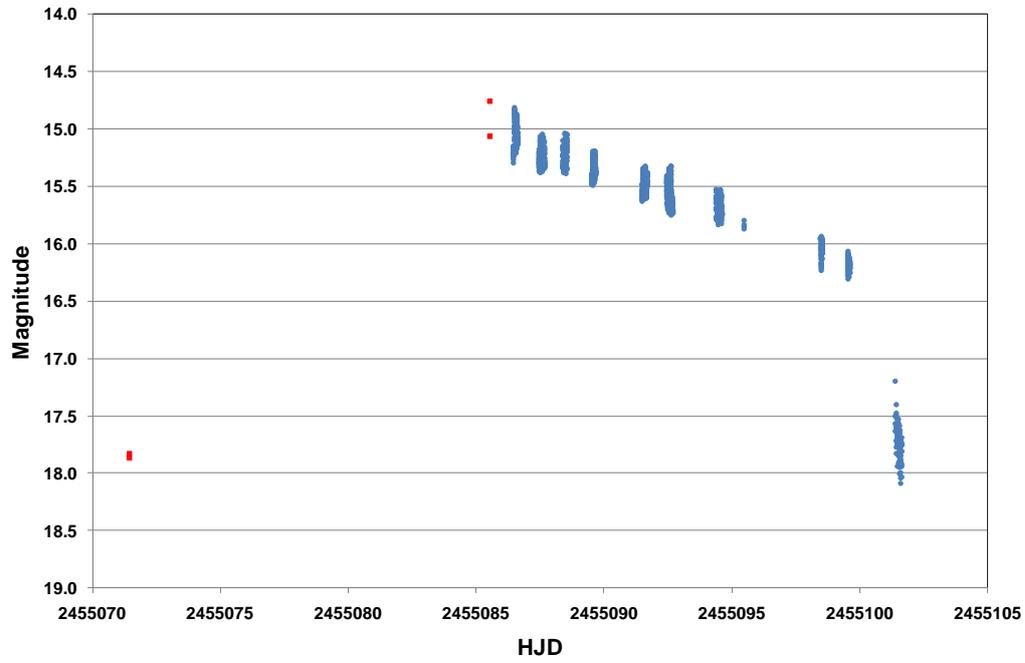

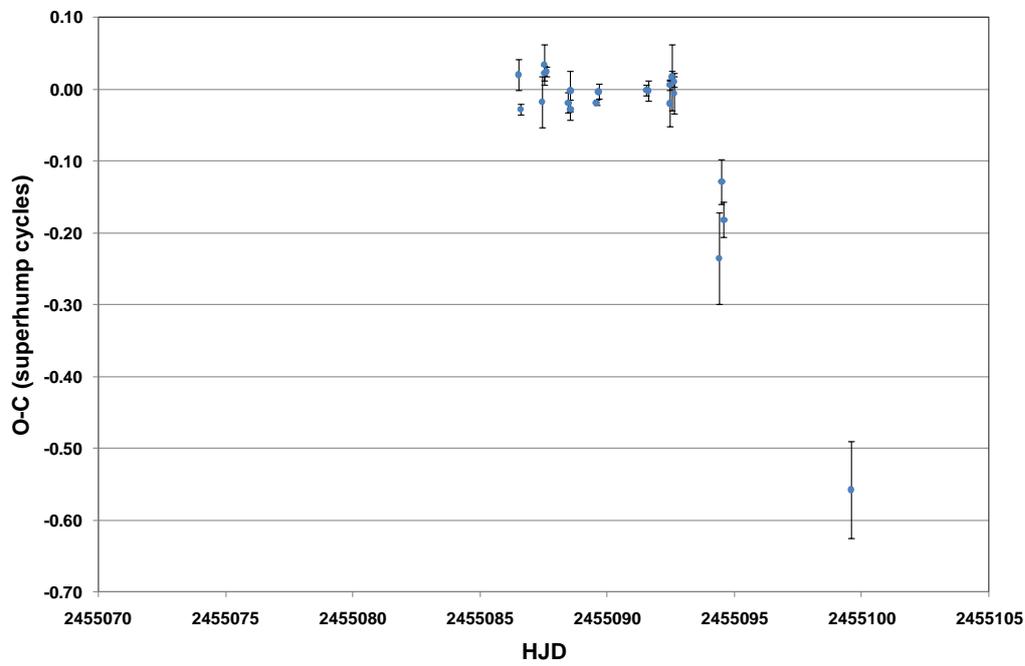

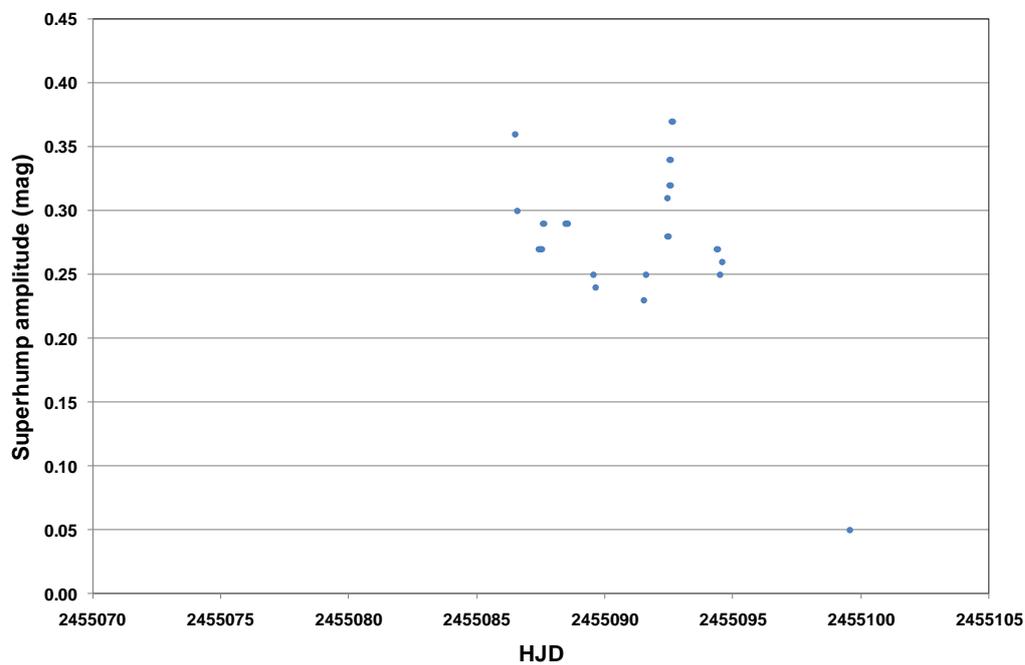

**Figure 1: the 2009 outburst of BG Ari. Top: outburst light curve. Middle: O-C diagram. Bottom: superhump amplitude. Red data points = data from CRTS, blue = data from the authors**

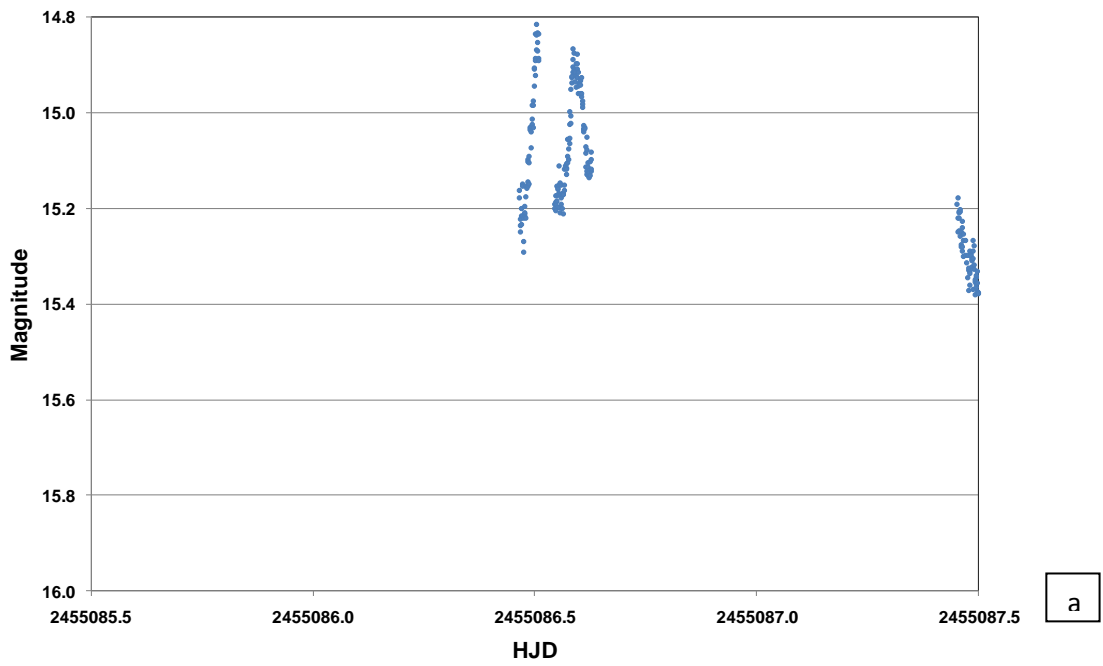

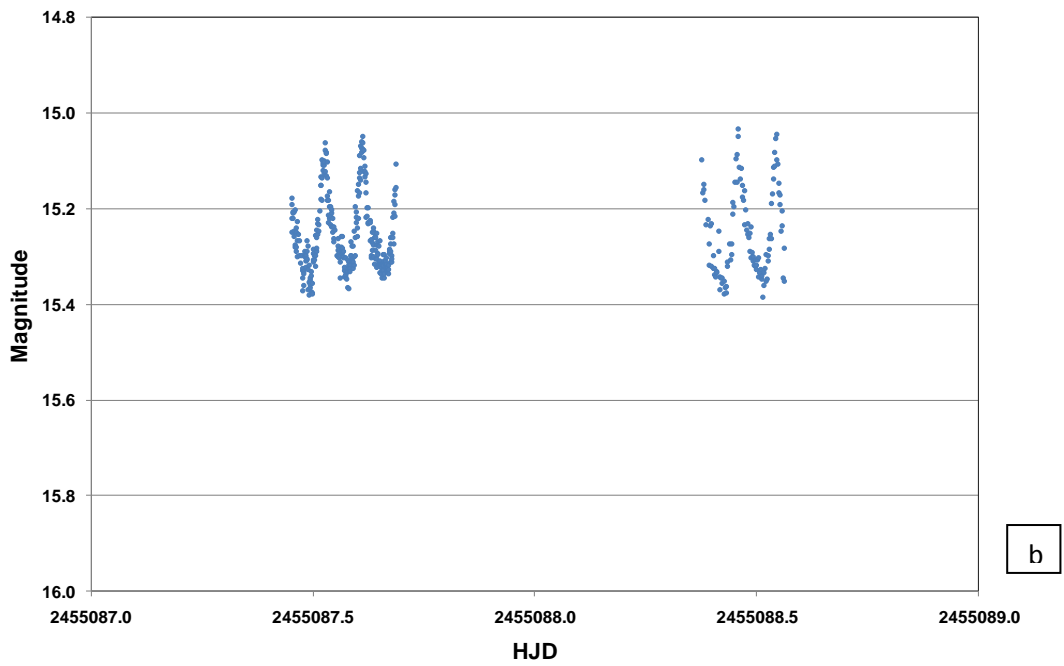

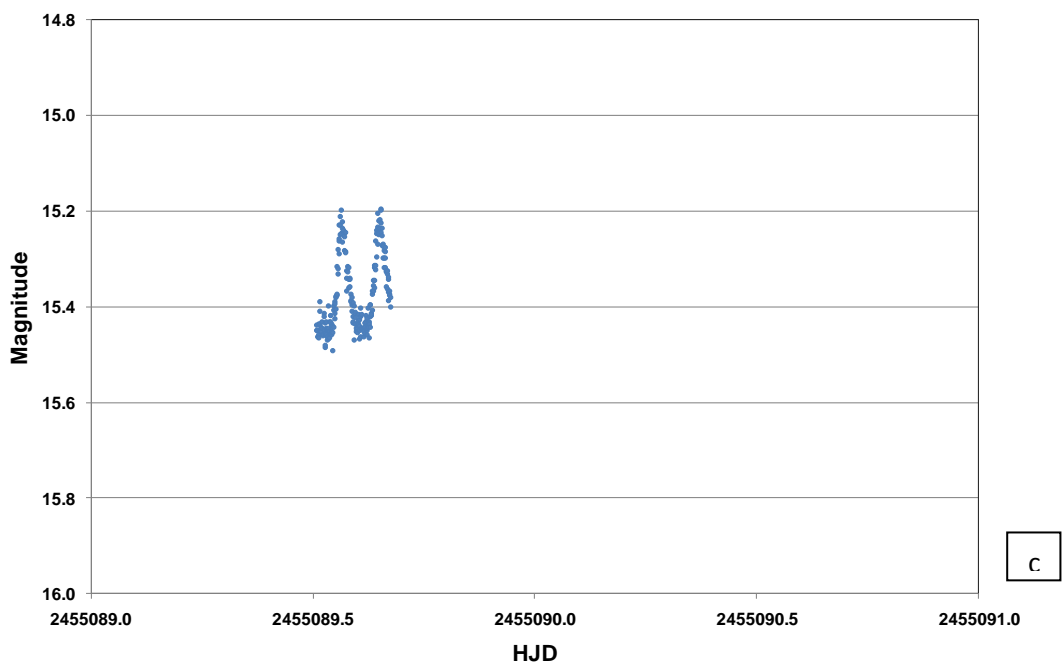

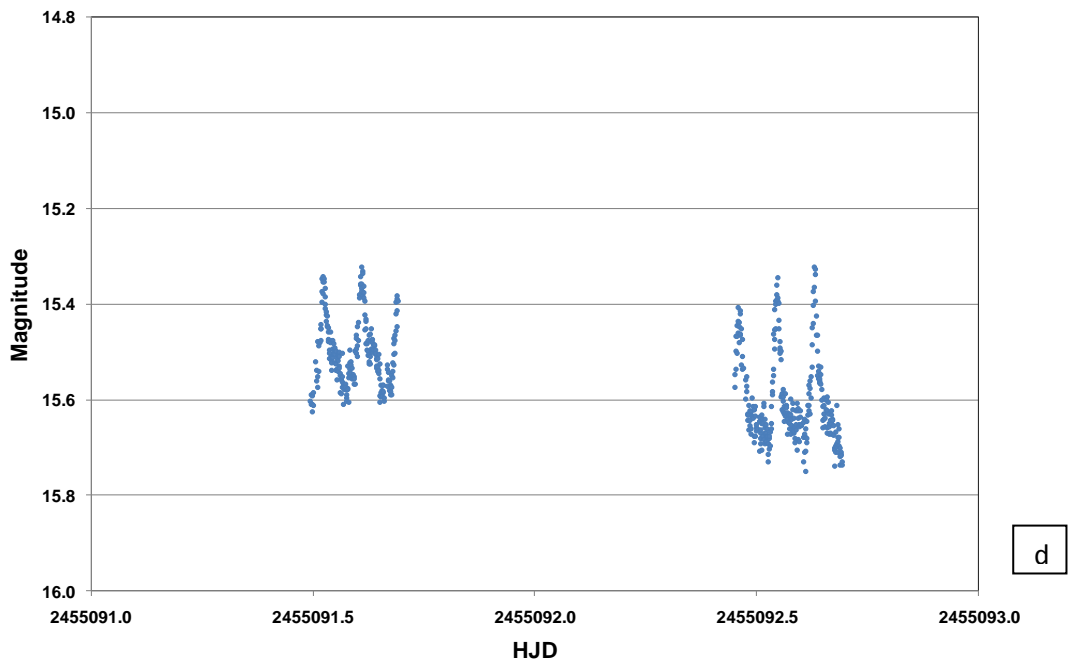

d

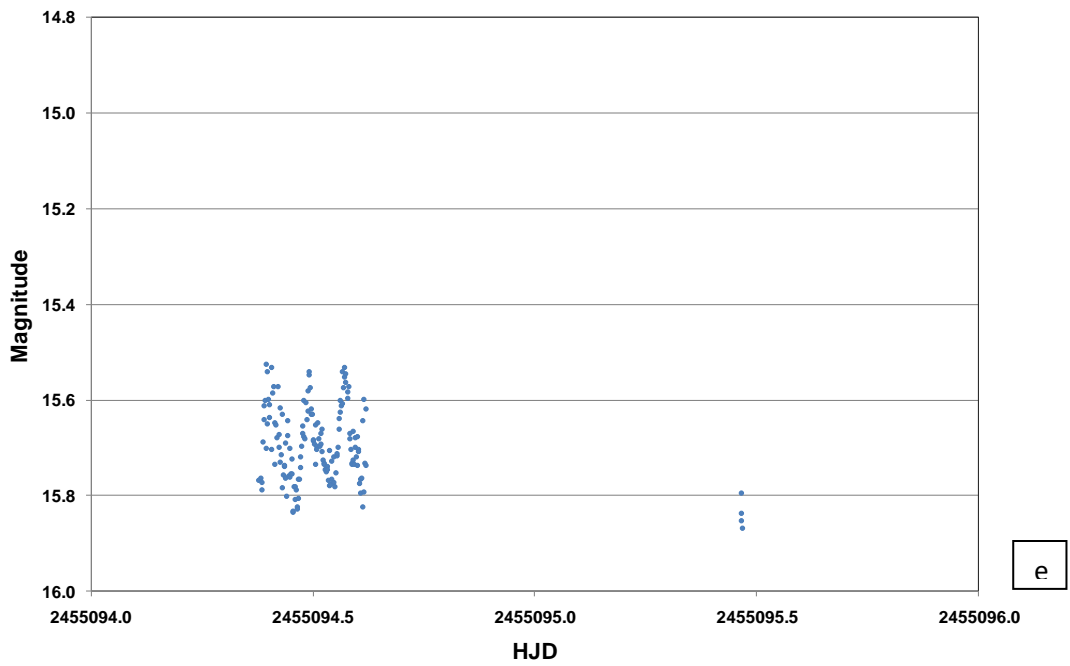

e

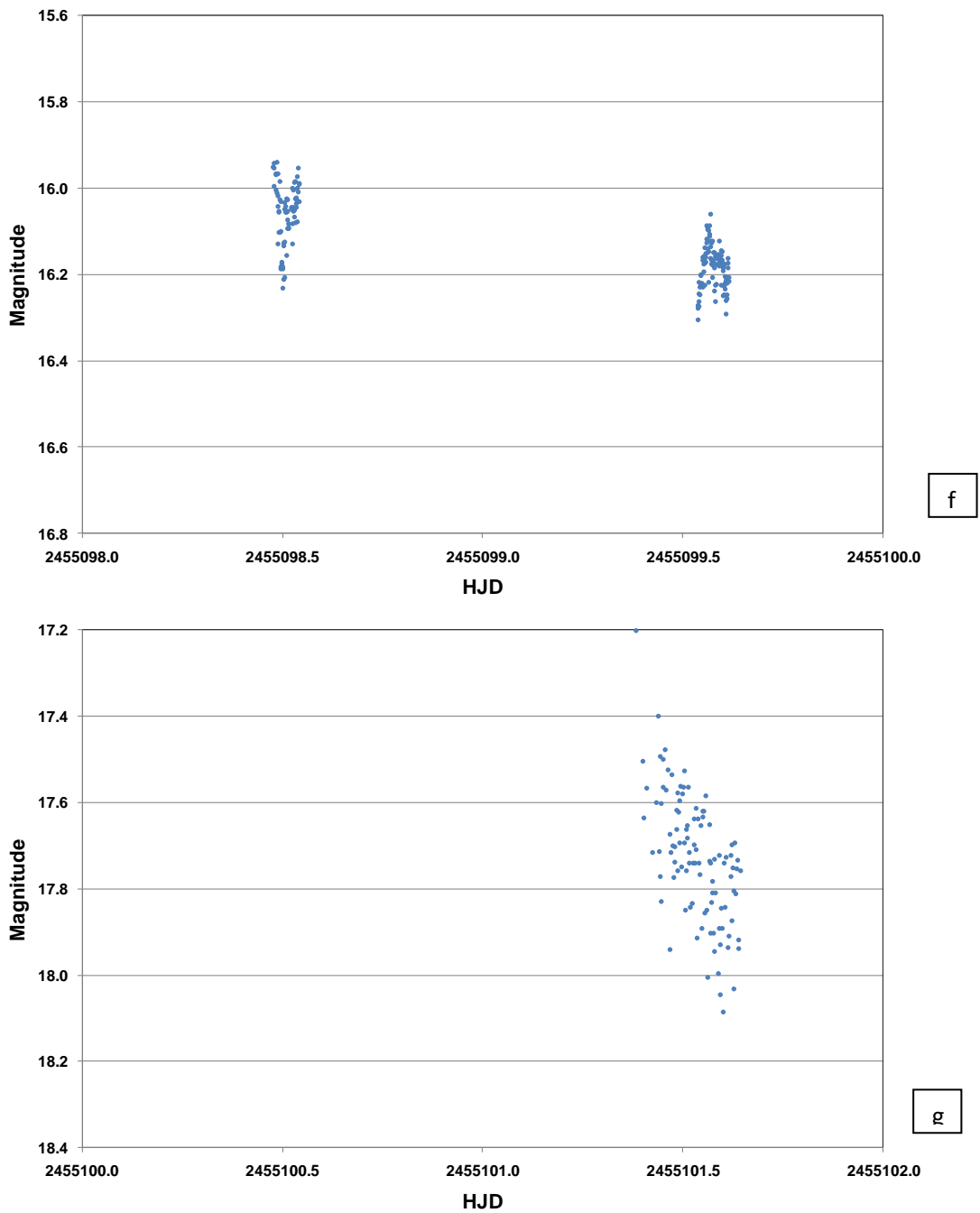

**Figure 2: Time resolved photometry during the 2009 outburst**

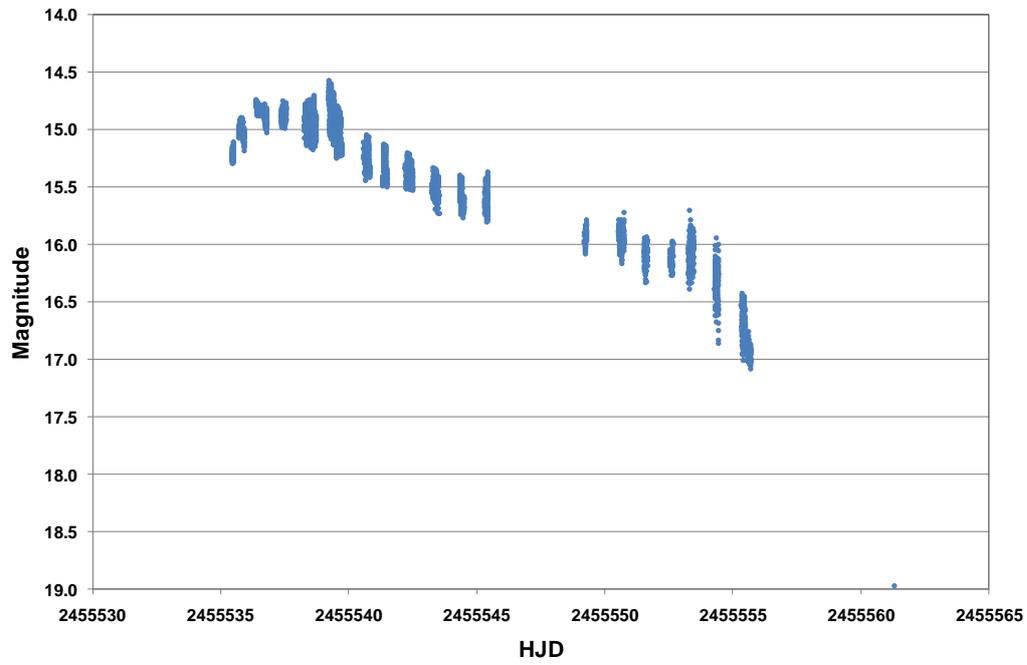

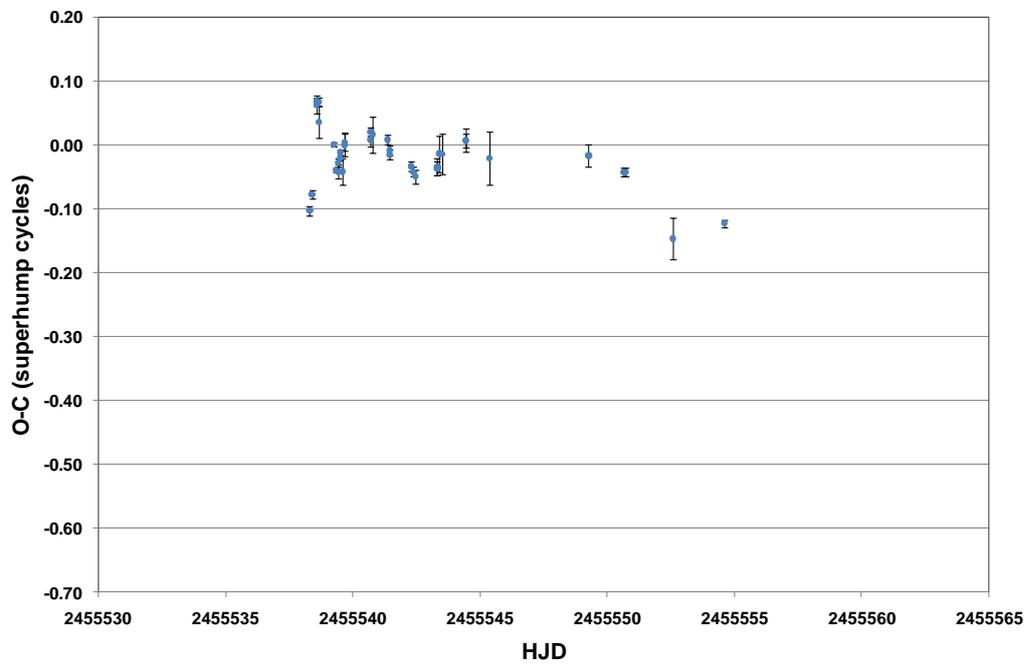

Figure 3: the 2010 outburst of BG Ari. Top: outburst light curve. Middle: O-C diagram. Bottom: superhump amplitude

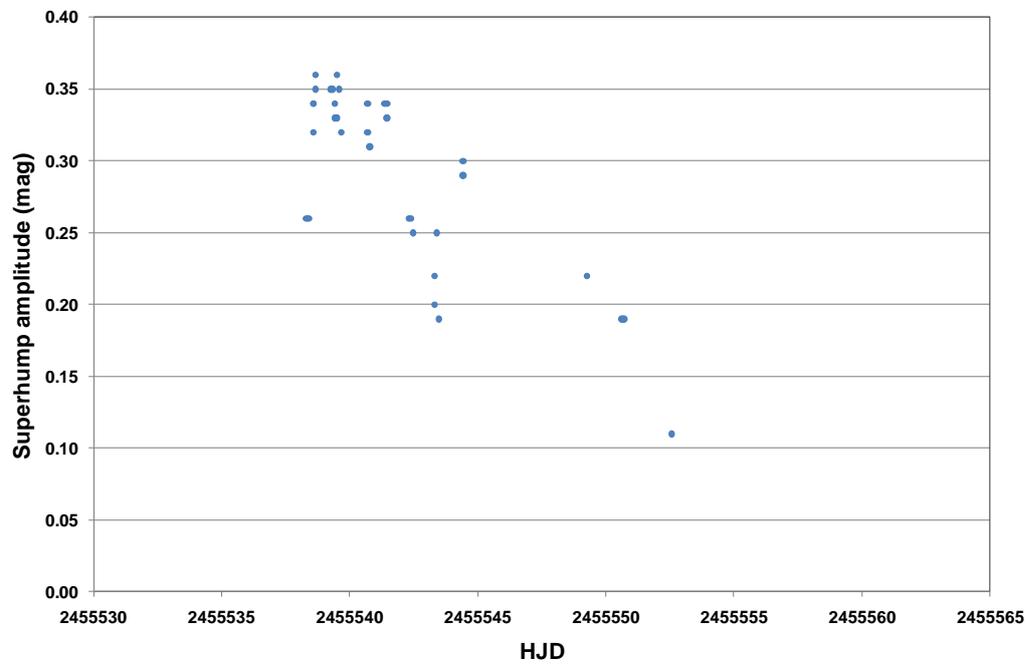

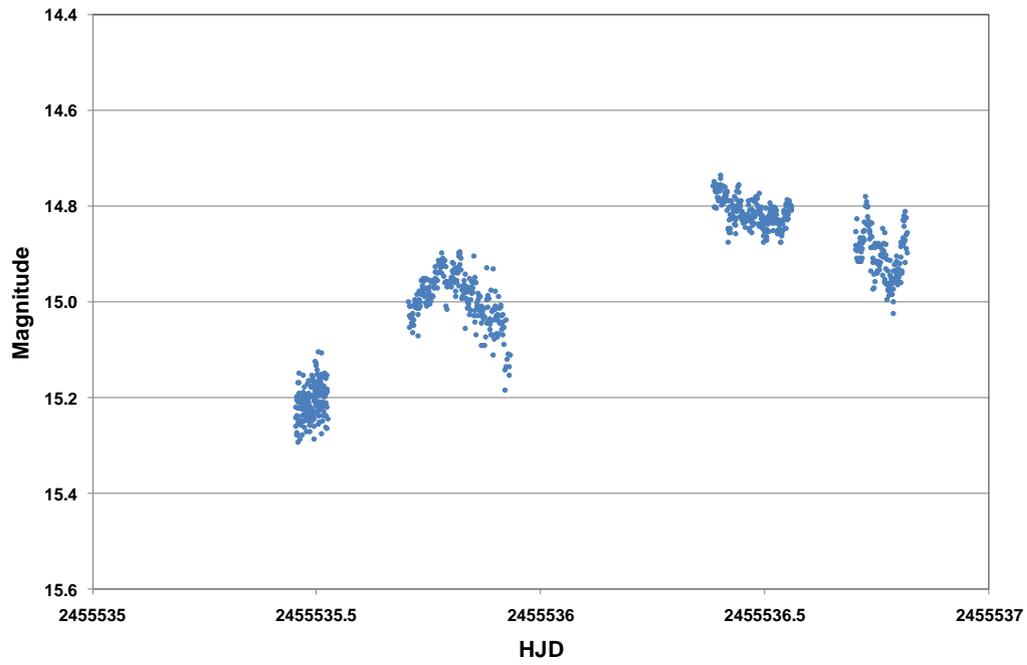

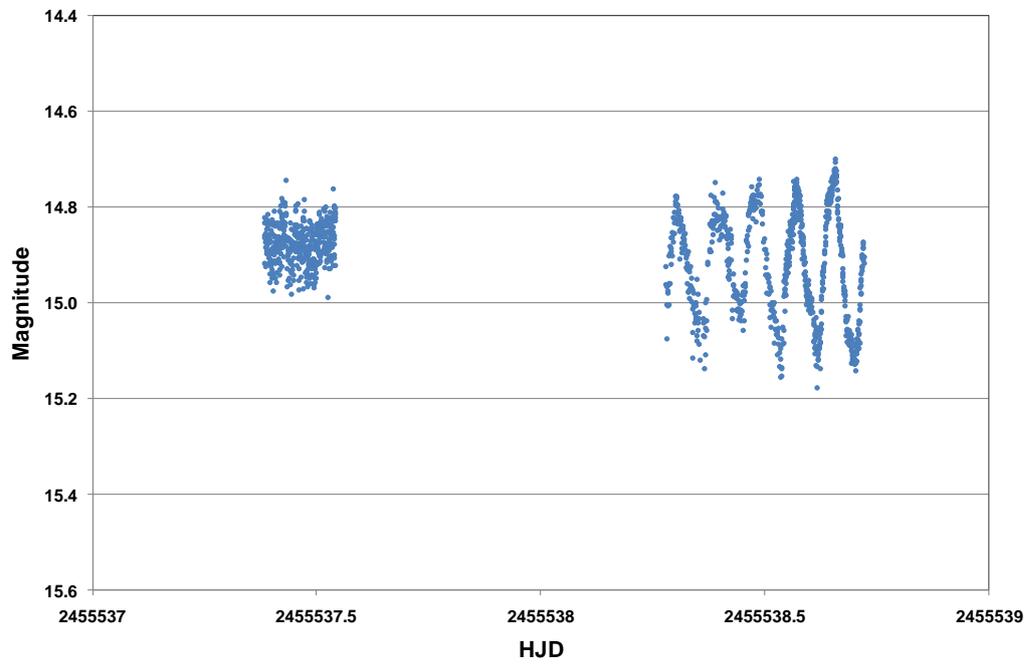

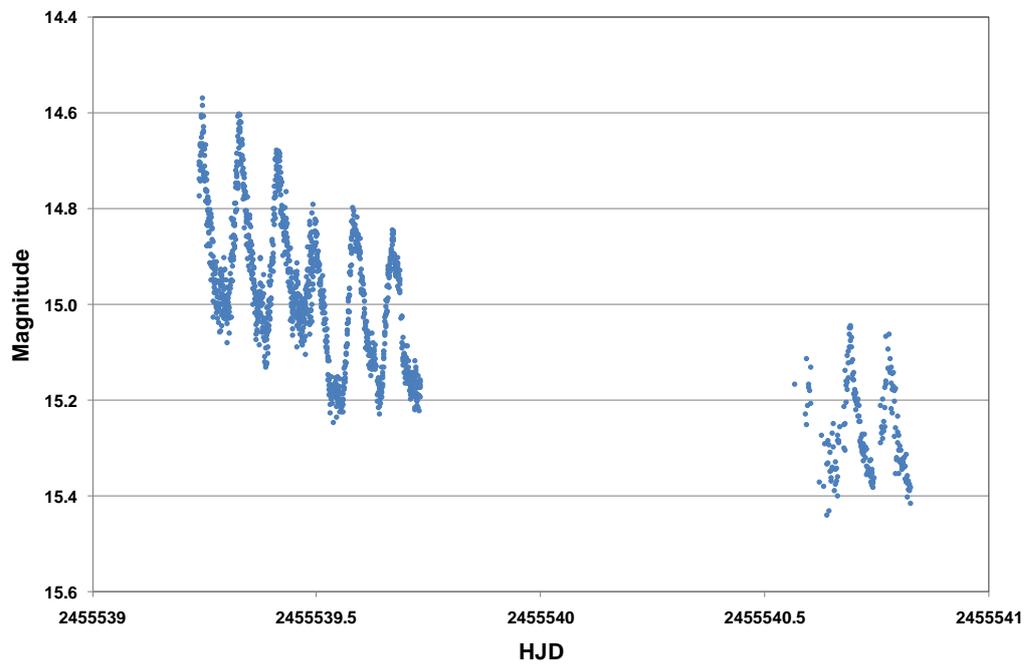

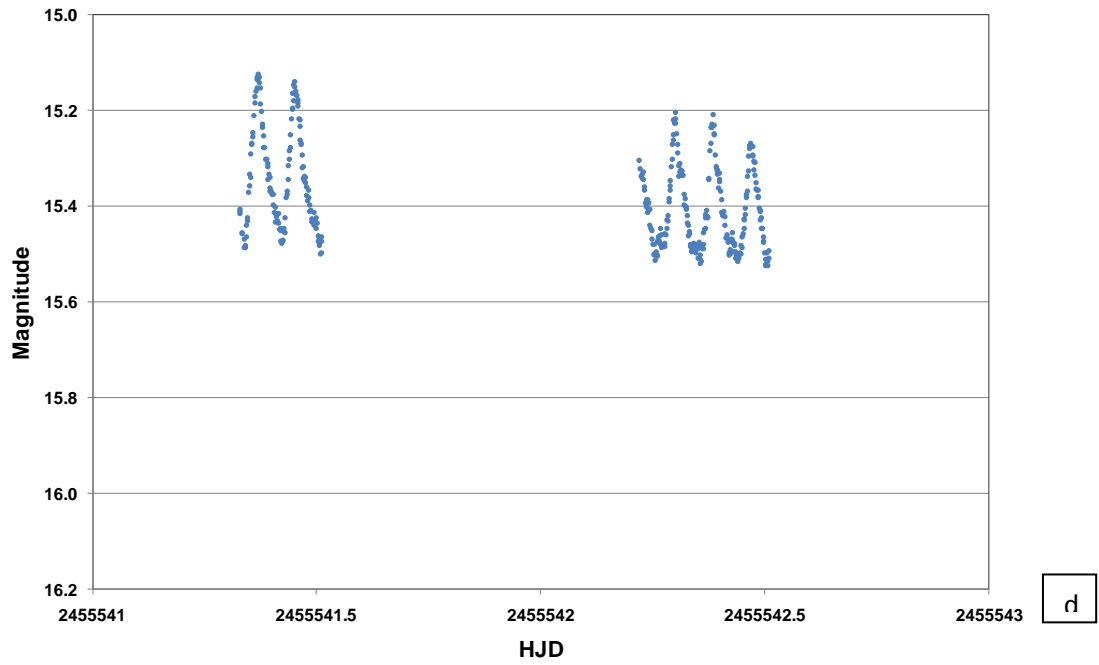

d

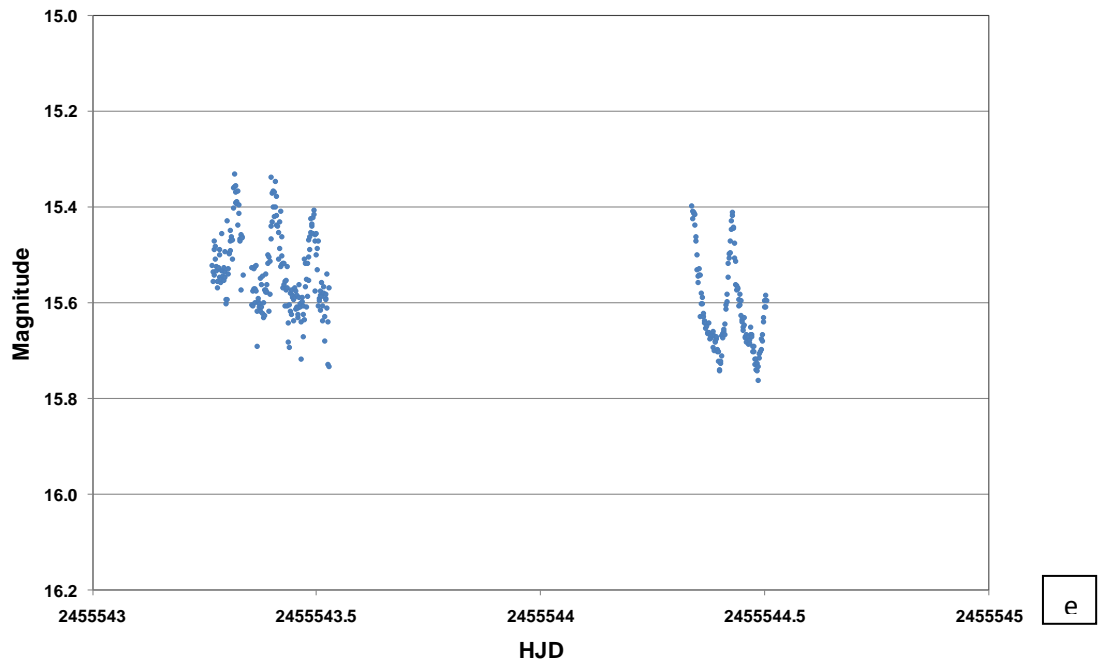

e

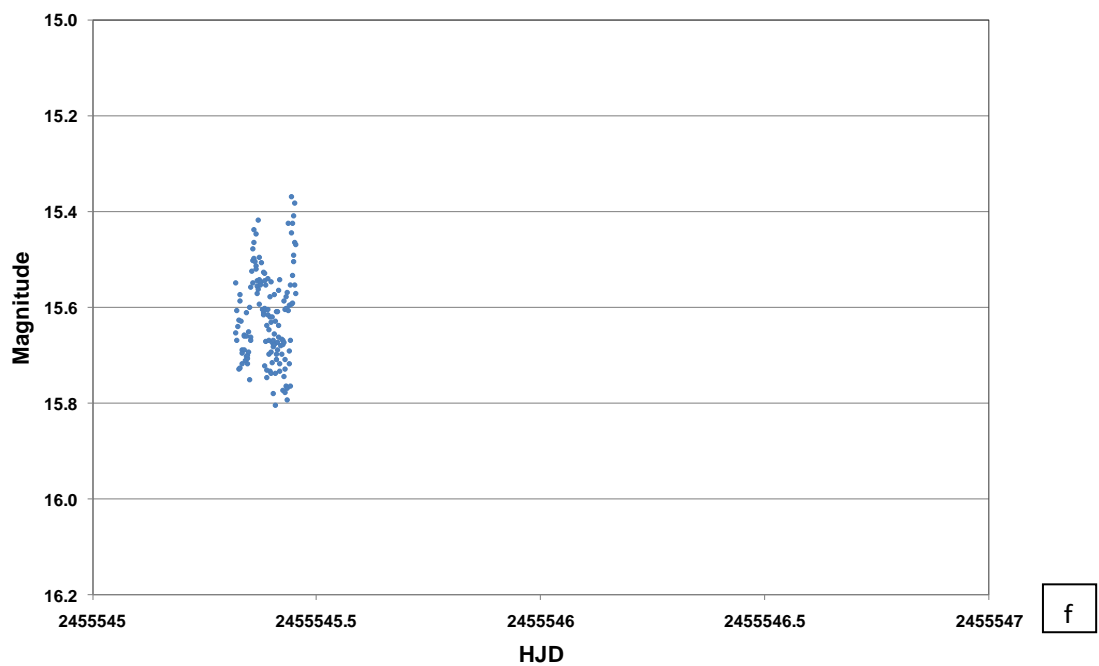

f

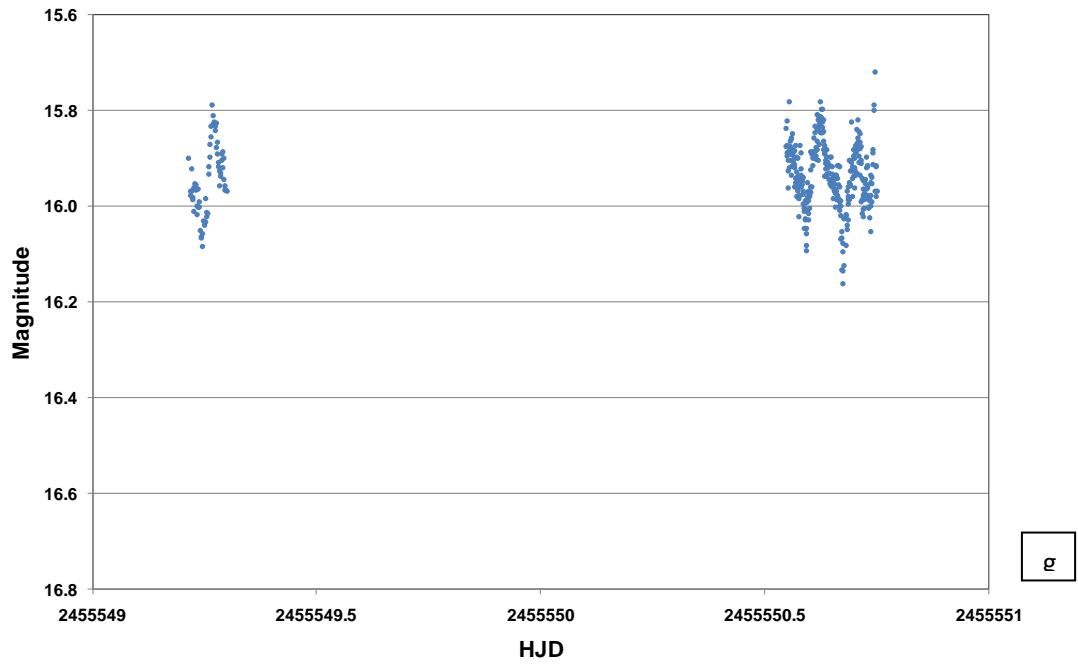

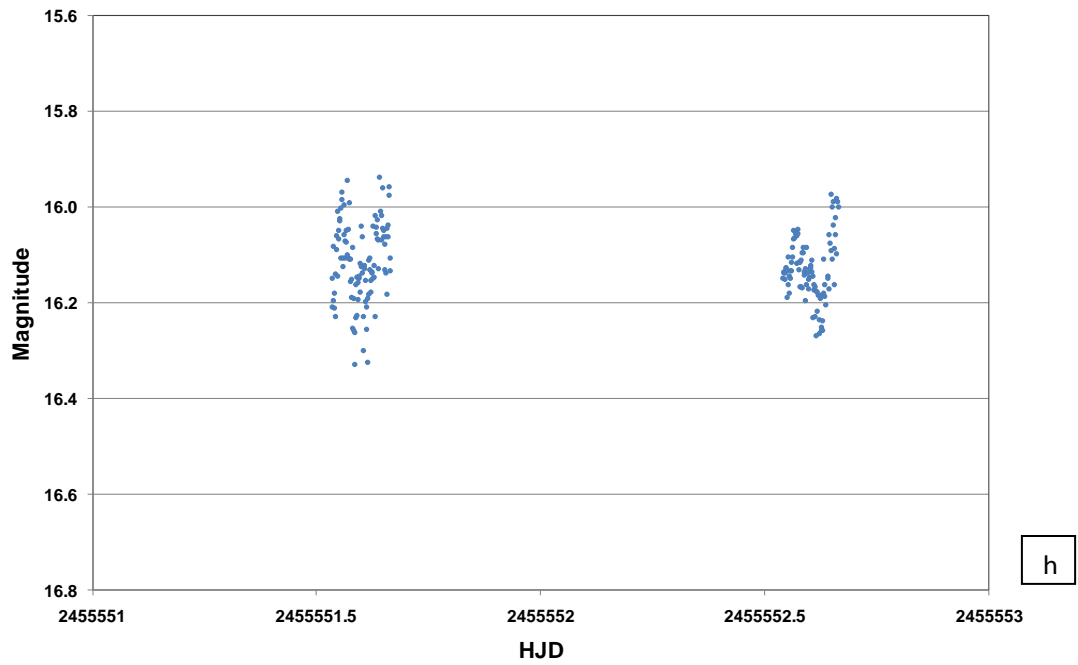

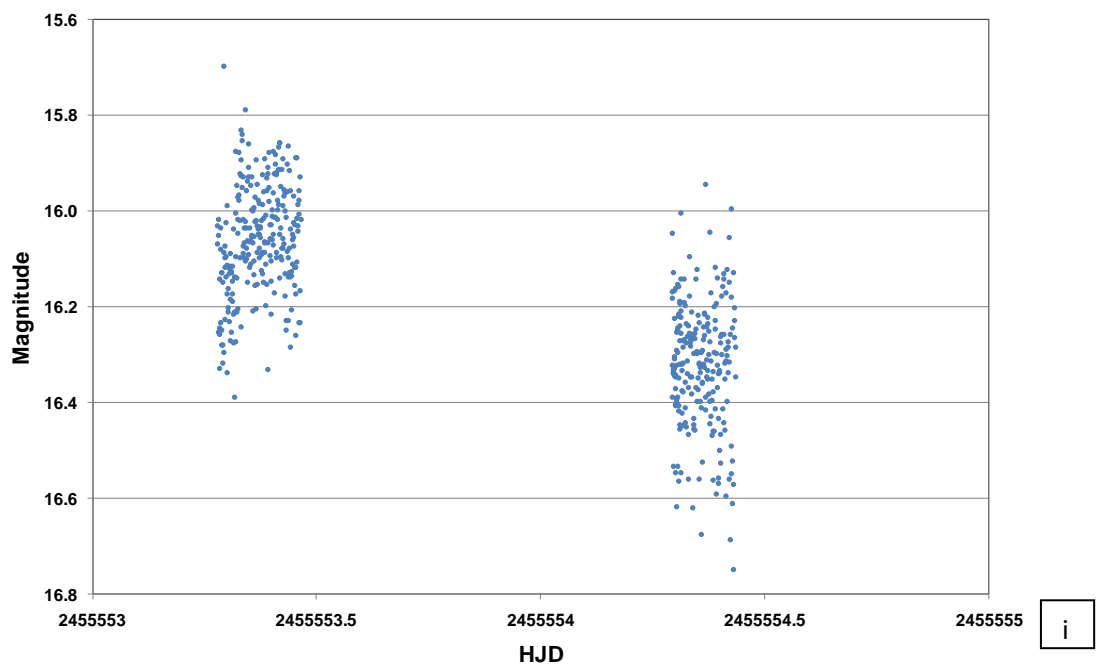

**Figure 4: Time resolved photometry during the 2010 outburst**

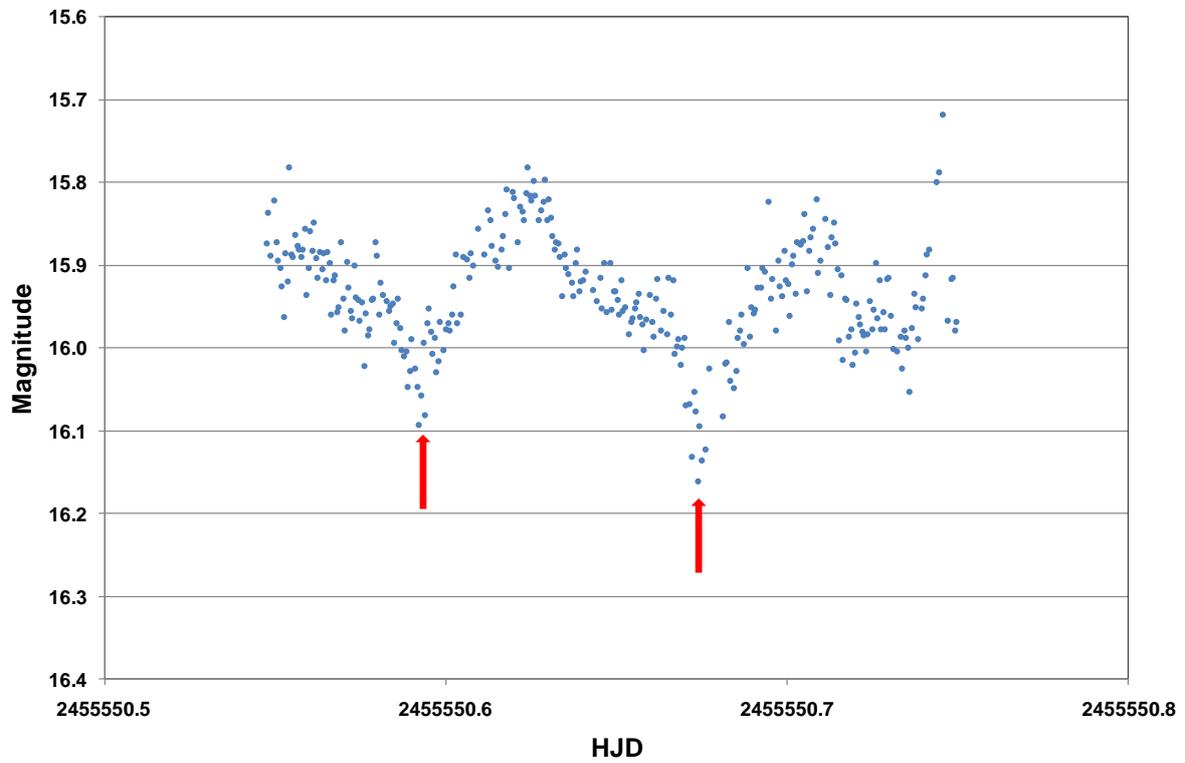

**Figure 5: Photometry from the 2010 outburst on HJD 245550 showing two eclipses**

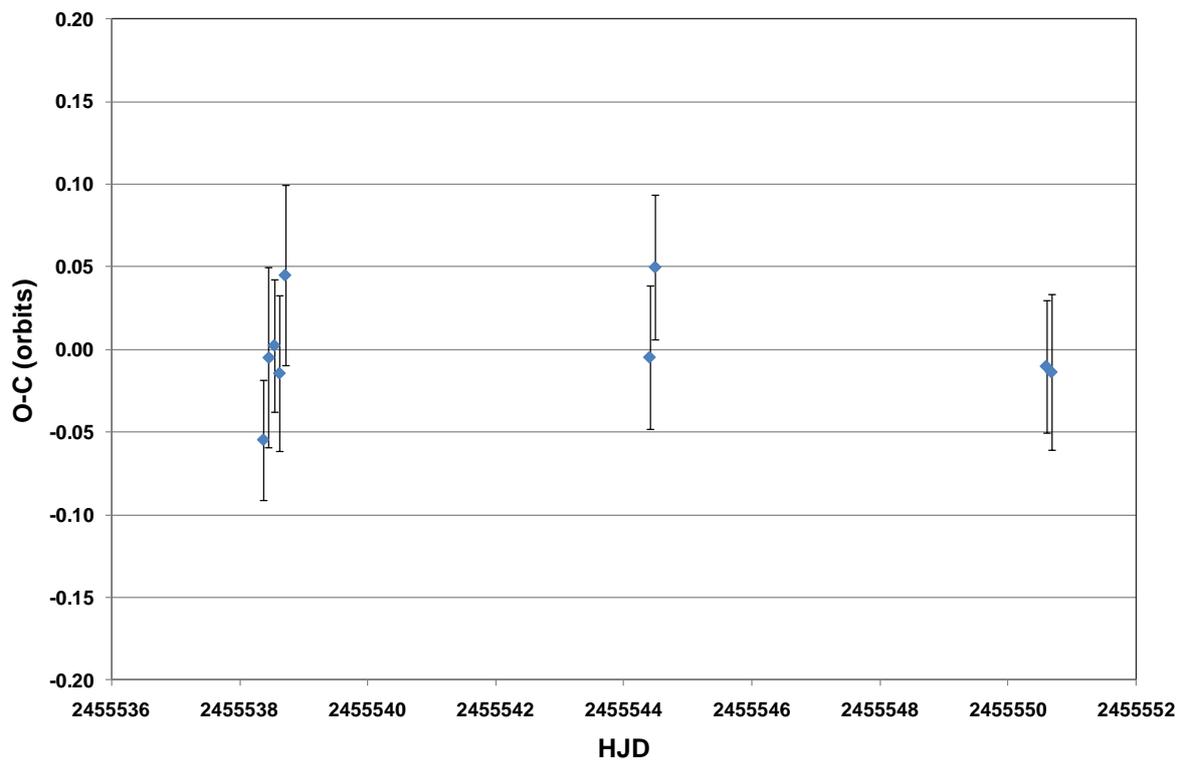

**Figure 6: O-C for the eclipses during the 2010 outburst**

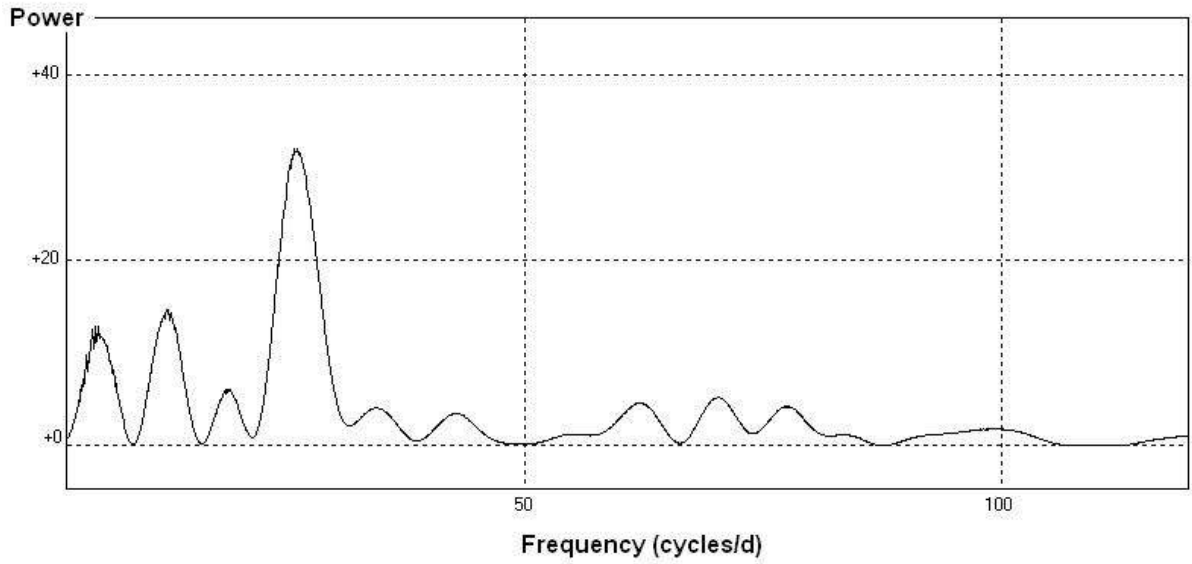

**(a)**

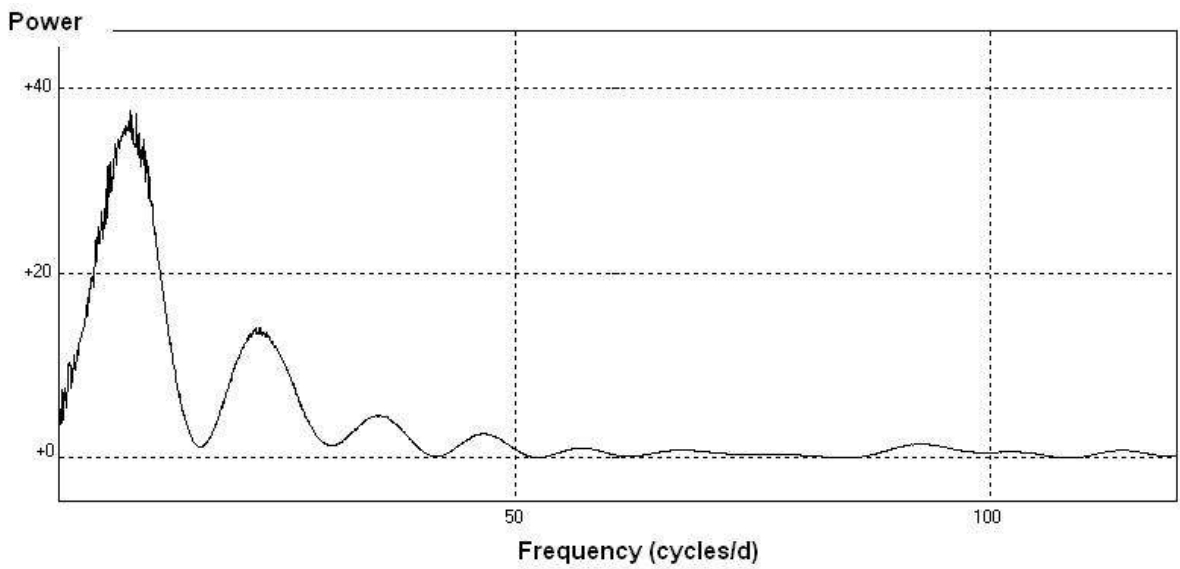

**(b)**

**Figure 7: Lomb-Scargle power spectra from the 2010 outburst**

(a) Data from HJD 2455536.3 and 2455536.6
(b) Data from HJD 2455536.7 and 2455536.9

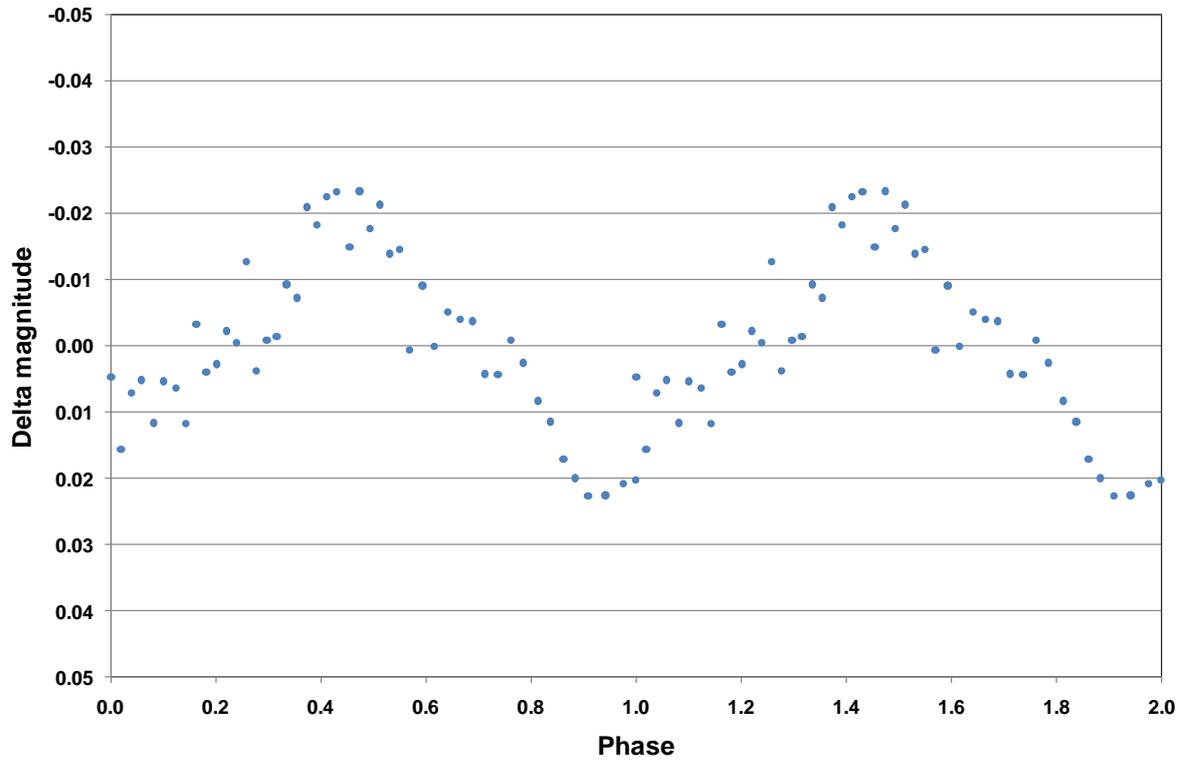

**Figure 8: Phase diagram of photometry from HJD 2455536.3 and 2455536.6 folded on P = 0.0389 d**